\documentclass[aps,pra,twocolumn,letterpaper,superscriptaddress,10pt]{revtex4-2}
\usepackage{amssymb,amsthm,amsmath,amsfonts}
\usepackage{graphicx,ulem,enumerate,bbm,bm,mathptmx}
\usepackage[pdftex,dvipsnames,usenames]{xcolor}
\usepackage[colorlinks=true,urlcolor=blue,citecolor=blue,linkcolor=blue]{hyperref}

\newcommand{\revision}[1]{\textcolor{black}{#1}}

\begin{document}

\title{%
    Entanglement between dependent degrees of freedom: Quasiparticle correlations
}

\author{Franziska Barkhausen}
    \email{fbarkhau@mail.uni-paderborn.de}
    \affiliation{Institute for Photonic Quantum Systems (PhoQS), Department of Physics and Center for Optoelectronics and Photonics Paderborn (CeOPP), Paderborn University, Warburger Strasse 100, D-33098 Paderborn, Germany}

\author{Laura Ares}
    \email{laura.ares.santos@uni-paderborn.de}
    \affiliation{Institute for Photonic Quantum Systems (PhoQS), Department of Physics and Center for Optoelectronics and Photonics Paderborn (CeOPP), Paderborn University, Warburger Strasse 100, D-33098 Paderborn, Germany}

\author{Stefan Schumacher}
    \affiliation{Institute for Photonic Quantum Systems (PhoQS), Department of Physics and Center for Optoelectronics and Photonics Paderborn (CeOPP), Paderborn University, Warburger Strasse 100, D-33098 Paderborn, Germany}
    \affiliation{Wyant College of Optical Sciences, University of Arizona, Tucson, Arizona 85721, USA}

\author{Jan Sperling}
    \affiliation{Institute for Photonic Quantum Systems (PhoQS), Department of Physics and Center for Optoelectronics and Photonics Paderborn (CeOPP), Paderborn University, Warburger Strasse 100, D-33098 Paderborn, Germany}

\date{\today}

\begin{abstract}

    Common notions of entanglement are based on well-separated subsystems.
    However, obtaining such independent degrees of freedom is not always possible because of physical constraints.
    In this work, we explore the notion of entanglement in the context of dependent degrees of freedom.
    As a physically relevant application, we specifically study quantum correlation features for quasiparticle descriptions.
    Those are paramount for interacting light-matter systems, utilizing excitations of fermion-boson hybrid modes.
    By comparing independent and dependent degrees of freedom, we uncover that certain states are nonentangled although they would be entangled when only focusing on the common, independent description, and vice versa.
    Therefore, insight is provided into the resourcefulness of quantum correlations within the rarely discussed context of dependent degrees of freedom for light-matter links in quantum information applications.

\end{abstract}

\maketitle


\section{Introduction}
\label{sec:Introduction}

    Features that surpass classical limitations are at the heart of quantum science and quantum technologies \cite{SAP17,CG19}.
    Arguably the most versatile quantum phenomenon is entanglement \cite{HHHH09}, being crucial for many applications in quantum communication and quantum computation \cite{NC00}, such as teleportation \cite{BBCJPW93} and dense coding \cite{BW92}.
    While determining whether a state is entangled is known to be a computationally hard problem \cite{I07}, certain states are easily recognizable as entangled, e.g., W states, GHZ states, and NOON states \cite{DVC00,S89,BKABWD00}.

    Entanglement in hybrid systems \cite{KL12,ANLF15} is crucial for quantum links between distinct physical platforms, e.g., using matter properties for processing and light for communication \cite{HSP10,ANLF15}.
    The quasiparticle description is a convenient way to describe such hybrid systems, including the assessment of quantum resources stemming from light-matter interactions  \cite{LPRSHSSA21,SSGR23}.
    Interfacing bosonic and fermionic parts in quantum links, quasiparticles thus constitute joint excitations of quantized bosonic and fermionic modes resulting in mixed quantum statistics for the quasiparticles \cite{MKMS24}.

    Various quantum properties of quasiparticles, including entanglement, have been studied \cite{RS98,LPRSHSSA21,PWAK23,CLMP23,CLFB23,SSGR23}.
    Still, the versatile nature of entanglement means that a plethora of different forms of entanglement exist \cite{HV13,LM13,SSV14,GSVCRT16}.
    For example, it can occur independently for particles and fields to the point that both forms can be classified as two distinct phenomena \cite{SPLAG23}.
    In addition, entanglement can be independent of the decomposition of the full system into quantized basis modes, allowing entanglement to persist regardless of the choice of a mode basis \cite{SPLBS19}.
    This and other findings raise the question of whether the common analysis of entanglement applies to the quasiparticle picture, too.

    In optics, for example, commonly studied degrees of freedom include, but are not limited to orbital angular momentum, time-bin modes, spatial modes, and polarization \cite{FT20}.
    In all these instances, entanglement is formulated in terms of independent degrees of freedom of purely bosonic excitations \cite{ZW16,XW18,FG20,ABEOBABUH23,LN24}.
    However, degrees of freedom can be dependent because of physical constraints, which is not commonly discussed and, therefore, is the focus of this work.

    In this work, we overcome the aforementioned limitation by characterizing entangled and nonentangled states between dependent degrees of freedom.
    This is mainly done for light-matter interfaces and their quasiparticle descriptions via a mode transformation to quasiparticle excitations arising from diagonalizing a Hamiltonian in terms of quantized fields.
    We show instances of both entangled and nonentangled states for dependent degrees of freedom whose independent counterparts behave oppositely, showing the distinctiveness of quantum correlations in such scenarios.
    Also, we compare the fermion-boson quasiparticle modes with analogous models consisting purely of bosonic and fermionic subsystems.
    Finally, a generalization to other scenarios and instances of  multipartite entanglement is outlined.

    The remainder of the work is structured as follows.
    \revision{The concepts of dependent and independent degrees of freedom in the context of quantum systems are established in Sec. \ref{sec:DefinitionSection}.}
    In Sec. \ref{sec:LightMatterInterface}, we introduce the specific light-matter system under study.
    Entanglement for quasiparticle descriptions is established and investigated in Sec. \ref{sec:QuasiParticleEntanglement}.
    A comparison with independent degrees of freedom, specifically, purely bosonic and purely fermionic scenarios, is discussed in Sec. \ref{sec:Comparison}.
    In Sec. \ref{sec:Generalization}, we discuss generalizations of our approach that apply to an arbitrary number of bosonic and fermionic modes.
    We conclude in Sec. \ref{sec:Conclusion}.


\section{Independent and dependent degrees of freedom for composite quantum systems}
\label{sec:DefinitionSection}

\revision{%
    The classification of quantum-physical degrees of freedom is motivated by the notion of statistical independence in probability theory.
    In this classical context, the ability to factorize a joint distribution of random variables into distributions for each variable is at the core of this concept.
    Based on the notion of statistical independence, a classification of dependent and independent degrees of freedoms can be introduced within quantum physics.
}

\revision{%
    For a composite quantum system, consisting of Hilbert spaces $\mathcal H_X$ and $\mathcal H_Y$, the principles of quantum physics for independent degrees of freedom require the joint system to read $\mathcal H_X\otimes\mathcal H_Y$.
    This space includes factorizable states $|x\rangle_X\otimes|y\rangle_Y$, as well as all linear combinations thereof.
    If this construction applies, we speak about independent degrees of freedom in which the common notion of entanglement is well defined;
    otherwise, we refer to $X$ and $Y$ as dependent.
}

\revision{%
    In the case of physical constraints limiting the available state space, one has to treat the scenario more carefully.
    This applies particularly if joint restrictions apply across the $X$ and $Y$ parts of the composite systems.
    For instance, the symmetrization postulate generally does not admit product states $|x\rangle_X\otimes|y\rangle_Y$ for identical particles $X$ and $Y$ in first quantization.
    Rather, the (anti)symmetrized states $|x\rangle_X\otimes|y\rangle_Y\pm|y\rangle_X\otimes|x\rangle_Y\pm$ form the basis of the state space for the now dependent degrees of freedom.
    Specifically note that symmetric and antisymmetric states are obtained via projections onto the symmetric and antisymmetric subspaces of the composite quantum system.
    This impacts the very definition of entanglement
    (see, e.g., Ref. \cite{RSV15} and the references therein).
}

    \revision{As we are going to explore in this work via a detailed study of quasiparticle excitations in second quantization, the entanglement characteristics of quasiparticles describe another example of dependent degrees of freedom.
    For instance, we show that nonentangled states are projected product states [see Eq. \eqref{eq:BBprojectedToFB}].
    Based on this result and the example mentioned above, we formalize this approach.}
    Let $\mathcal H=\hat P(\mathcal H_X\otimes\mathcal H_Y)$ be a composite Hilbert space subject to a constraint given by the projection $\hat P$, with $\hat P^2=\hat P=\hat P^\dag$.
    A pure nonentangled state can be defined as
    \begin{equation}
        \label{eq:GenDef}
        |x,y\rangle_{XY}=\hat P\left(
            |x\rangle_X \otimes |y\rangle_Y
        \right),
    \end{equation}
    where $|x\rangle_X\in\mathcal H_X$ and $|y\rangle_Y\in\mathcal H_Y$ are local states of the joint system.
    Note that this definition is identical with the one for a factorizable state for independent degrees of freedom, where $\hat P=\hat 1$.
    An entangled state $|\Psi\rangle$ does not admit the decomposition from Eq. \eqref{eq:GenDef}, $|\Psi\rangle\neq \hat P\,  (|x\rangle_X \otimes |y\rangle_Y)$. 
    Of course, we choose $\langle\Psi|\Psi\rangle=1$ such that the state is normalized after projection.

    Furthermore, a mixed, only classically correlated state takes the form
    \begin{equation}
    	\hat\rho_\mathrm{sep.}
    	=\hat P\left(
            \int dP(x,y)
            |x\rangle_{X\,X}\langle x|
            \otimes
            |y\rangle_{Y\,Y}\langle y|
    	\right)\hat P^\dag,
    \end{equation}
    and a state that does not obey this form is consequently inseparable, i.e., entangled.
    Also note that this definition straightforwardly generalizes to multipartite systems, \revision{and it is compatible with the well-established notion of separability and inseparability for independent degrees of freedom \cite{W89} when $\hat P=\hat 1$ holds true}.
    Therefore, a separable state is here defined as a state that obeys the physical constraint (given by the projector $\hat P$) and only contains classical correlations (via the joint probability distribution $P$) of tensor-product states $|x\rangle_X \otimes |y\rangle_Y$.

    \revision{In addition to the quasiparticle description we explore in the remainder of this work, we draw a comparison to fundamental observables.}
    Consider the degree of freedom pertaining to the total angular momentum, $J^2$, and one for the angular momentum in the $z$ direction, $J_z$.
    It is well known that the corresponding angular-momentum operators can be jointly diagonalized,
    \begin{equation}
    \begin{aligned}
    	\hat J^2|j,m\rangle_{J^2J_z}
    	={}&\hbar^2j(j+1)|j,m\rangle_{J^2J_z}
    	\\
    	\text{and}\quad
    	\hat J_z|j,m\rangle_{J^2J_z}
    	={}&\hbar m|j,m\rangle_{J^2J_z},
    	\\
    	\text{with}\quad
    	-j\leq m\leq j
    	&\quad\text{and}\quad
    	j\geq0.
    \end{aligned}
    \end{equation}
    From the latter restriction of values, the imposed physical constraint can be reformulated through the projection
    \begin{equation}
    	\hat P
    	=
    	\sum_{j:j\geq0}|j\rangle_{J^2\,J^2}\langle j|
    	\otimes
    	\sum_{m:-j\leq m\leq j}
    	|m\rangle_{J_z\,J_z}\langle m|,
    \end{equation}
    where $j$ and $m$ could, in principle, take arbitrary (half) integer values without the restrictions.
    The projection $\hat P$ then makes sure that the angular momentum in one direction never exceeds the total angular momentum, $|m|\ngtr j$.
    The above definition further allows us to define nonentangled states as follows:
    \begin{equation}
    	|x,y\rangle_{J^2\,J_z}=\hat P
    	\left(
            \sum_{j}x_j|j\rangle_{J^2}
        \right)
        \otimes
    	\left(
            \sum_{m}y_{m}|j\rangle_{J_z}
        \right).
    \end{equation}
    For instance, the state $(|0,0\rangle_{J^2\,J_z}+|1,-1\rangle_{J^2\,J_z}+|1,0\rangle_{J^2\,J_z}+|1,1\rangle_{J^2\,J_z})/2$ is not of the form $|x\rangle_{J^2}\otimes |y\rangle_{J_z}$ itself, but it can be obtained via a projection of the local states $|x\rangle_{J^2}\propto |0\rangle_{J^2}+|1\rangle_{J^2}$ and $|y\rangle_{J_z}\propto |-1\rangle_{J_z}+|0\rangle_{J_z}+|-1\rangle_{J_z}$, in which the projection removes the unphysical parts $|0\rangle_{J^2}\otimes|\pm1\rangle_{J_z}$.
    See also Ref. \cite{APALBSS24} for the relation to other notions of nonclassicality.


\revision{
    For the remainder of this work, we compare entanglement for dependent and independent degrees of freedom in the context of light-matter interfaces.
    Those systems are of particular interest from the physical as well as from the quantum-informational perspective, as outlined in the Introduction.
}


\section{Light-matter interface}
\label{sec:LightMatterInterface}

    We begin by introducing the specific system under study.
    This particular system functions as a well-studied, proof-of-concept example for analyzing entanglement of independent and dependent degrees of freedom throughout this work.

\subsection{Quantized eigenmodes of a Hamiltonian}

\begin{figure}
	\includegraphics[width=.75\columnwidth]{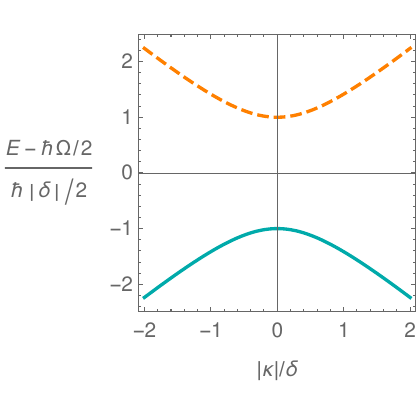}
	\caption{%
        Energy bands $E_\pm$ from diagonalization of the Hamiltonian via the introduction of quasiparticle modes $\hat p_\pm$ in Eq. \eqref{eq:QuasiparticleBands} as a function of the ratio of the coupling strength $|\kappa|$ and the detuning $\delta$.
	}\label{fig:Bands}
\end{figure}

    Consider a composite fermion-boson system in second quantization with independent field operators $\hat f$ and $\hat b$, respectively, obeying $\{\hat f,\hat f\}=0=[\hat b,\hat b]$ and $\{\hat f,\hat f^\dag\}=\hat 1=[\hat b,\hat b^\dag]$, where $\hat 1$ denotes the identity.
    The system's Hamiltonian under study consists of free-field contributions and couplings, analogous to the Jaynes--Cummings model \cite{JC63}, reading as
    \begin{equation}
      \label{eq:HamiltonianFB}
        \frac{\hat H}{\hbar}
        =\omega_F\hat f^\dag\hat f
        +\omega_B\hat b^\dag\hat b
        +\frac{\kappa}{2}\hat f^\dag\hat b
        +\frac{\kappa^\ast}{2}\hat b^\dag\hat f,
    \end{equation}
    with a complex coupling parameter $\kappa\neq0$.
    It is convenient to introduce the abbreviations
    \begin{equation}
    	\Omega=\omega_F+\omega_B
    	\quad\text{and}\quad
    	\delta=\omega_F-\omega_B,
    \end{equation}
    where $\delta$ is dubbed the detuning.
    Using an $\mathrm{SU}(2)$ map, we can find the eigenmodes of the Hamiltonian, here named $\hat p_\pm$, as
    \begin{equation}
        \label{eq:ModeTransformation}
    	\begin{bmatrix}
    		\hat p_+
    		\\
    		\hat p_-
    	\end{bmatrix}
    	=\begin{bmatrix}
            \varphi & \beta e^{i\vartheta}
            \\
            -\beta e^{-i\vartheta} & \varphi
        \end{bmatrix}
    	\begin{bmatrix}
    		\hat f
    		\\
    		\hat b
    	\end{bmatrix},
    \end{equation}
    where
    \begin{equation}
        \label{eq:Parameters}
    	\varphi=\sqrt{\frac{\Delta+\delta}{2\Delta}},
    	\quad
    	\beta=\sqrt{\frac{\Delta-\delta}{2\Delta}},
    	\quad\text{and}\quad
    	\frac{\kappa}{|\kappa|}=e^{i\vartheta},
    \end{equation}
    with $\Delta=\sqrt{\delta^2+|\kappa|^2}$ and $\varphi^2+\beta^2=1$.
    Excitations of these eigenmodes define quasiparticles, which we later discuss in detail.
    See also Ref. \cite{BD24} in this context.
    Furthermore, we can recast the Hamiltonian in Eq. \eqref{eq:HamiltonianFB} in a noninteracting form as
    \begin{equation}
        \label{eq:QuasiparticleBands}
    	\hat H
    	=E_+\hat p_+^\dag\hat p_+
    	+E_-\hat p_-^\dag\hat p_-,
    	\quad\text{with}\quad
    	E_\pm=\frac{\hbar}{2}\left(\Omega\pm\Delta\right)
    \end{equation}
    and without cross-terms between the quasiparticle modes;
    see also Fig. \ref{fig:Bands} for the energies pertaining to the eigenmodes.

\subsection{Quantum states of the system}

    Beginning with the vacuum state $|\mathrm{vac}\rangle$, the Fock basis in the boson-fermion description consists of the elements
    \begin{equation}
        \label{eq:BasisStatesFB}
    	|0,n\rangle_\mathrm{FB}=\frac{\hat b^{\dag n}}{\sqrt{n!}}|\mathrm{vac}\rangle
    	\quad\text{and}\quad
    	|1,n\rangle_\mathrm{FB}=\hat f^\dag\frac{\hat b^{\dag n}}{\sqrt{n!}}|\mathrm{vac}\rangle,
    \end{equation}
    for $n\in\mathbb N$ and utilizing $\hat f^{\dag m}=0$ for all $m>1$.
    In addition to the vacuum state, all eigenstates (also known as dressed states \cite{VW06}) of the Hamiltonian can be expanded as
    \begin{equation}
        \label{eq:Eigenstates}
    \begin{aligned}
        |\phi_{+,N}\rangle=&{}
        \beta_N|0,N\rangle_\mathrm{FB}
        +\varphi_Ne^{i\vartheta}|1,N-1\rangle_\mathrm{FB}
        \\
        \text{and}\quad
        |\phi_{-,N}\rangle=&{}
        \varphi_N|0,N\rangle_\mathrm{FB}
        -\beta_Ne^{i\vartheta}|1,N-1\rangle_\mathrm{FB},
    \end{aligned}
    \end{equation}
    with Rabi frequencies $\Delta_N=\sqrt{\delta^2+N|\kappa|^2}$ and parameters
    \begin{equation}
        \label{eq:ParametersIndexedN}
        \varphi_N=\sqrt{\frac{\Delta_N+\delta}{2\Delta_N}}
        \quad\text{and}\quad
        \beta_N=\sqrt{\frac{\Delta_N-\delta}{2\Delta_N}}.
    \end{equation}
    Note that, for the single-particle excitation, $N=1$, we have $\Delta_1=\Delta$, $\varphi_1=\varphi$, and $\beta_1=\beta$, according to our definitions in Eq. \eqref{eq:Parameters}.
    Moreover, the energies pertaining to the eigenstates in Eq. \eqref{eq:Eigenstates} are
    \begin{equation}
        \label{eq:EnergySpectrum}
    	E_{\pm,N}=\frac{\hbar}{2}\left(N\Omega-N\delta+\delta\pm\Delta_N\right).
    \end{equation}

    The notion of entanglement in the original $|m,n\rangle_\mathrm{FB}$ is given in terms of independent degrees of freedom.
    However, we are going to explore entanglement with respect to the eigenmodes of the Hamiltonian, in which we separate quasiparticles excited through $\hat p_+$ from the excitations of $\hat p_-$, resulting in dependent degrees of freedom.


\section{Quasiparticle entanglement}
\label{sec:QuasiParticleEntanglement}

    In this section, we analyze the notion of entanglement for dependent degrees of freedom for the model under study.
    We show how linear dependence plays a central role in this concept.
    Also, we explore how the different notions of entanglement for states of a given total particle number behave differently for independent and dependent degrees of freedom.

\subsection{Product states in fermion-boson hybrid descriptions}

    For defining entanglement of dependent degrees of freedom, we first need to recapitulate nonentangled states for independent degrees of freedom.
    Specifically, product states of independent fermions and bosons take the form
    \begin{equation}
        \label{eq:classicalformFB}
    \begin{aligned}
    	|x,y\rangle_\mathrm{FB}
    	=&{}
    	\left(
            \sum_{m=0}^1 x_m|m\rangle_\mathrm{F}
    	\right)
    	\otimes
    	\left(
            \sum_{n=0}^\infty y_m|m\rangle_\mathrm{B}
    	\right)
    	\\
    	=&{}
    	\sum_{m=0}^1\sum_{n=0}^\infty
    	\tilde x_m\tilde y_n
    	\hat f^{\dag m}\hat b^{\dag n}
    	|\mathrm{vac}\rangle,
    \end{aligned}
    \end{equation}
    where $\tilde x_m=x_m$ and $\tilde y_n=y_n/\sqrt{n!}$ relate expansion coefficients in the Fock basis and excitations of quantized fields.
    As examples, the basis states in Eq. \eqref{eq:BasisStatesFB} are classical states with respect to the fermion-boson separation.
    A state $|\psi\rangle$ that cannot be expressed as defined in Eq. \eqref{eq:classicalformFB} is, by definition, entangled.

    Note that, throughout most parts of this work, we focus on pure states.
    However, it is worth mentioning that classically correlated states, i.e., separable states \cite{W89}, can be introduced by using an ensemble of pure product states in the form of a density matrix.
    Still, all concepts discussed here can be extended to mixed, inseparable states, as discussed in Sec. \ref{sec:DefinitionSection}.

\subsection{Field-product states for dependent degrees of freedom}

    Generalizing the second line in Eq. \eqref{eq:classicalformFB} renders it possible to define nonentangled states for the quasiparticle modes $\hat p_\pm$, Eq. \eqref{eq:ModeTransformation}.
    That is, a state is said to be nonentangled in this $\pm$ description if it can be expressed in a factorized manner as follows:
    \begin{equation}
        \label{eq:classicalformPM}
    	|x,y\rangle_\pm
    	=
    	\left(\sum_{m\in\mathbb N} \tilde x_m\hat p_+^{\dag m}\right)
    	\left(\sum_{n\in\mathbb N} \tilde y_n\hat p_-^{\dag n}\right)
    	|\mathrm{vac}\rangle,
    \end{equation}
    with complex coefficients $x_m$ and $y_n$ such that the state is normalized, ${}_\pm\langle x,y|x,y\rangle_\pm=1$.
    \revision{In the next section, we show that this operational definition exactly coincides with the one for dependent degrees from Sec. \ref{sec:DefinitionSection}.}

    According to the definition in Eq. \eqref{eq:classicalformPM}, the following $N$-particle states, where $N=m+n$, are product states:
    \begin{equation}
    \begin{aligned}
    	&{}\hat p_+^{\dag m}\hat p_-^{\dag n}|\mathrm{vac}\rangle
    	\\
    	=&{}
    	e^{-im\vartheta}\left(
            \beta^m\hat b^{\dag m}
            +m\varphi\beta^{m-1}e^{i\vartheta}\hat f^\dag\hat b^{\dag[m-1]}
    	\right)
    	\\
    	&{}
    	\times\left(
            \varphi^n\hat b^n
            -n\beta\varphi^{n-1}e^{i\vartheta}\hat f^\dag\hat b^{\dag[n-1]}
    	\right)|\mathrm{vac}\rangle
    	\\
    	=&e^{-im\vartheta}\beta^{m}\varphi^{n}\sqrt{N!}
    	\\
    	&{}\times
        \left(
            |0,N\rangle_\mathrm{FB}
            +\frac{m\varphi^2-n\beta^2}{\beta\varphi\sqrt{N}}e^{i\vartheta}|1,N-1\rangle_\mathrm{FB}
        \right),
    \end{aligned}
    \end{equation}
    being also expanded in the $\mathrm{FB}$ basis for comparison.
    Including a proper normalization and ignoring global phases, this yields the $\pm$-product states
    \begin{equation}
        \label{eq:GeneratingStates}
        |m,n\rangle_\pm
        =
        \frac{
            |0,N\rangle_\mathrm{FB}
            +\frac{m\varphi^2-n\beta^2}{\beta\varphi\sqrt{N}}e^{i\vartheta}|1,N-1\rangle_\mathrm{FB}
        }{\sqrt{
            1
            +\frac{(m\varphi^2-n\beta^2)^2}{\beta^2\varphi^2N}
        }}.
    \end{equation}
    Note that such states are generally product states with respect to the quasiparticle modes $\hat p_\pm$ but entangled with respect to the fermion-boson separation $\mathrm{FB}$ since a global mode transformation is used in Eq. \eqref{eq:ModeTransformation}.
    Also, we can choose $m,n\in\mathbb N$ since neither $\hat p_+^{\dag m}$ nor $\hat p_-^{\dag n}$ becomes zero for a finite integer, contrasting pure fermions.

\subsection{Linear dependence and disentangled NOON states}

    It is straightforward to observe that, for any two pairs of integers $(m,n)$ and $(m',n')$, which obey $(m\varphi^2-n\beta^2)/\sqrt{m+n}\neq (m'\varphi^2-n'\beta^2)/\sqrt{m'+n'}$, the states $|m,n\rangle_\pm$ and $|m',n'\rangle_\pm$ are linearly independent.
    However, for a fixed $N>0$, there are only two linearly independent elements in total since the $\mathrm{FB}$ expansion is based on two basis elements only [see Eq. \eqref{eq:GeneratingStates}].
    Therefore, the identities
    \begin{equation}
    \begin{aligned}
        |1,N-1\rangle_\mathrm{FB}
        =&{}
        \frac{e^{-i\vartheta}\varphi\beta}{\sqrt N}
        \left(
        \sqrt{1+N\frac{\varphi^2}{\beta^2}}|N,0\rangle_\pm
        \right.
        \\
        &{}
        \left.
        -\sqrt{1+N\frac{\beta^2}{\varphi^2}}|0,N\rangle_\pm
        \right)
    \end{aligned}
    \end{equation}
    and
    \begin{equation}
    \begin{aligned}
        |0,N\rangle_\mathrm{FB}
        =&{}
        \varphi^2\sqrt{1+N\frac{\beta^2}{\varphi^2}}|0,N\rangle_\pm
        \\
        &{}
        +\beta^2\sqrt{1+N\frac{\varphi^2}{\beta^2}}|N,0\rangle_\pm
    \end{aligned}
    \end{equation}
    imply for all $m=N-n$ and $n\notin\{0,N\}$ that we find
    \begin{equation}
         \label{eq:lincombinationBF}
    \begin{aligned}
    	|m,n\rangle_\pm
    	=&{}
    	\frac{m}{N}
    	\sqrt{\frac{
            N\varphi^2\beta^2+N^2\varphi^4
    	}{
            N\varphi^2\beta^2
            +\left(m\varphi^2-n\beta^2\right)^2
    	}}
    	|N,0\rangle_\pm
    	\\
    	&{}+
    	\frac{n}{N}
    	\sqrt{\frac{
            N\varphi^2\beta^2+N^2\beta^4
    	}{
            N\varphi^2\beta^2
            +\left(m\varphi^2-n\beta^2\right)^2
    	}}
    	|0,N\rangle_\pm.
    \end{aligned}
    \end{equation}
    For $N>2$, this shows that all but two $\pm$-product states can be expressed via superpositions of only two elements.

\begin{figure}
	\includegraphics[width=.8\columnwidth]{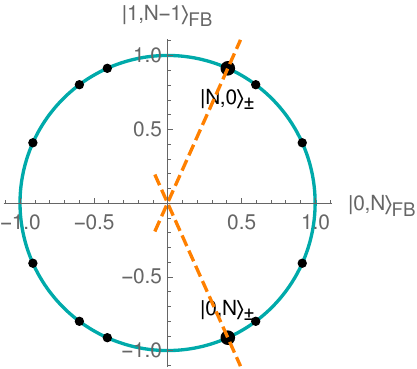}
	\caption{%
        NOON-type states from Eq. \eqref{eq:genNOON} on the circle in an FB coordinate system, for $N=5$ quasiparticles, where $\varphi=\beta=1/\sqrt2$ and $\vartheta=0$.
        All nonentangled states in the $\pm$ description, Eq. \eqref{eq:lincombinationBF}, are shown as bullet points, including a $\pm$ coordinate system of the dependent degrees of freedom, which is indicated through dashed lines.
        Note that $N+1=6$ states, plus their reflections about the origin, are shown in the real plane, ignoring complex phases of $\psi_0$ and $\psi_N$ for simplicity.
        For independent degrees of freedom, all but the states $|N,0\rangle_\pm$ and $|0,N\rangle_\pm$ (and their reflections $-|N,0\rangle_\pm$ and $-|0,N\rangle_\pm$) alone would be entangled.
	}\label{fig:NOON}
\end{figure}

    As a consequence of the above linear combination, so-called NOON states \cite{S89,BKABWD00},
    \begin{equation}
        \label{eq:genNOON}
    	|N00N\rangle_\pm=\psi_0|N,0\rangle_\pm+\psi_N|0,N\rangle_\pm,
    \end{equation}
    with $\psi_0\psi_N\neq0$, are not necessarily entangled for $N>2$.
    Recall that, by contrast, all NOON states are entangled for independent degrees of freedom.
    Here, whenever the coefficients $\psi_0$ and $\psi_N$ take the form of Eq. \eqref{eq:lincombinationBF}, such NOON states clearly admit a product form $|m,n\rangle_\pm$ for the dependent degrees of freedom $\pm$.
    Again, we emphasize that this disentanglement of quasiparticle NOON states is distinct from common discussions of entanglement (see Fig. \ref{fig:NOON}).

\subsection{Quantum-correlated eigenstates of noninteracting Hamiltonians}

    Next, we consider the noninteracting case, where $\kappa=0$ for the Hamiltonian in Eq. \eqref{eq:HamiltonianFB}, and the independent degrees of freedom $\mathrm{FB}$.
    This yields a Hamiltonian which decomposes into the single-mode parts $\hbar\omega_F\hat f^\dag\hat f$ and $\hbar\omega_B\hat b^\dag\hat b$, as well as into vanishing interactions, $\hat H_\mathrm{int}=0$.
    In that case, the eigenstates are necessarily products, $|m,n\rangle_\mathrm{FB}$ [Eq. \eqref{eq:BasisStatesFB}], for the nondegenerate case $\delta\neq0$.
    This is generally true for independent degrees of freedom and $\hat H=\hat H_1\otimes\hat 1_2+\hat 1_1\otimes\hat H_2$, with subsystem identities $\hat 1_j$ and local Hamiltonians $\hat H_j$ for $j\in\{1,2\}$.
    Further, the quasiparticle modes $\hat p_\pm$ were introduced to achieve a noninteracting representation of the Hamiltonian for $\kappa\neq0$ [Eq. \eqref{eq:QuasiparticleBands}].
    So the expectation derived from independent degrees of freedom would be that the $N$-particle eigenstates in Eq. \eqref{eq:Eigenstates} are products with respect to the $\pm$ separation.

    By equating coefficients of the eigenstates of $\hat H$ and the $N$-particle product states in Eq. \eqref{eq:GeneratingStates}, we find
    the following two possible conditions for factorization:
    \begin{equation}
        \label{eq:ConditionsFactEigen}
    	\frac{\varphi_N}{\beta_N}=\frac{m\varphi^2-n\beta^2}{\varphi\beta\sqrt{N}}
    	\quad\text{or}\quad
    	-\frac{\beta_N}{\varphi_N}=\frac{m\varphi^2-n\beta^2}{\varphi\beta\sqrt{N}},
    \end{equation}
    while recalling $\varphi_1=\varphi$, $\beta_1=\beta$, $N=m+n$, and Eq. \eqref{eq:ParametersIndexedN}.
    For $N>1$, those conditions are generally not met, therefore demonstrating that the eigenstates of an interaction-free Hamiltonian in dependent degrees of freedom can, in fact, be entangled.
    For example, the case $\delta=0$ applied to Eq. \eqref{eq:ConditionsFactEigen} simplifies the conditions therein to
    \begin{equation}
    	m=\frac{N\pm\sqrt{N}}{2}
    	\quad\text{and}\quad
    	n=\frac{N\mp\sqrt{N}}{2},
    \end{equation}
    which have integer solutions if and only if the total particle number $N$ is a perfect square.
    Conversely, $\sqrt{N}\notin\mathbb N$ leads to entangled $N$-quasiparticle eigenstates, contrary to the separation of eigenstates for independent degrees of freedom without coupling terms in the Hamiltonian.

\subsection{Discussion}

    We established the notion of classically correlated states for dependent degrees of freedom, being based on products states in Eq. \eqref{eq:classicalformPM}.
    Formally, those states admit the same structure as nonentangled states for independent degrees of freedom [Eq. \eqref{eq:classicalformFB}].
    However, we proved by examples two key distinguishing features of the different kinds of degrees of freedom.

    In the first example, we showed linear dependence in the dependent case.
    This led to the observation of disentangled NOON states, which are always entangled for independent degrees of freedom.
    The second example pertained to the product nature of eigenstates of noninteracting Hamiltonians, which are products for independent degrees of freedom.
    Again, a departure from this behavior was observed for dependent degrees of freedom.

    Therefore, quantum correlations of dependent degrees of freedom defy common assumptions about quantum correlations in a given system.
    This is interesting, e.g., in the context of light-matter interfaces for quantum information applications which require a firm understanding of quantum correlations for applications in various protocols.


\section{Comparing different quantum statistics}
\label{sec:Comparison}

    Thus far, we compared the separation with respect to independent degrees of freedom, consisting of one fermionic and one bosonic field (i.e., $\mathrm{FB}$ separation), and the quasiparticle description via hybrid eigenmodes (i.e., a separation with respect to the $+$ and $-$ modes), obeying neither a fermionic nor a bosonic algebra.
    In this section, we directly compare the previous cases with composite systems consisting purely of fermionic and bosonic quantized fields.

\subsection{Fermions, bosons, and combinations thereof}

    For the sake of simplicity, we assume a vanishing detuning, $\delta=0$ (likewise, $\omega_F=\omega=\omega_B$), and a zero phase for the coupling parameter, $\kappa/|\kappa|=e^{i\theta}=1$, throughout this section.
    For example, this yields balanced superposition [see Eq. \eqref{eq:ModeTransformation}] for the quasiparticle modes
    \begin{equation}
    	\hat p_\pm=\frac{\hat f\pm\hat b}{\sqrt2}.
    \end{equation}
    The factorizable $N$-quasiparticle states from Eq. \eqref{eq:GeneratingStates} simplify to
    \begin{equation}
    \label{eq:sepquasi}
    	|m,n\rangle_\pm=\frac{|0,N\rangle_\mathrm{FB}+\frac{m-n}{\sqrt{N}}|1,N-1\rangle_\mathrm{FB}}{\sqrt{1+\frac{(m-n)^2}{N}}},
    \end{equation}
    where $N=m+n$.
    Note that the vacuum state is trivially the same for all representations, $|\mathrm{vac}\rangle=|0,0\rangle_\mathrm{FB}=|0,0\rangle_\pm$.

    Analogously to the previous scenario, Eq. \eqref{eq:HamiltonianFB}, we can consider Hamiltonians
    $\hat H/\hbar=\omega(\hat f_1^\dag\hat f_1+\hat f_2^\dag\hat f_2)+|\kappa|(\hat f_1^\dag\hat f_2+\hat f_2^\dag\hat f_1)/2$
    and
    $\hat H/\hbar=\omega(\hat b_1^\dag\hat b_1+\hat b_2^\dag\hat b_2)+|\kappa|(\hat b_1^\dag\hat b_2+\hat b_2^\dag\hat b_1)/2$
    for purely fermionic and bosonic systems in second quantization, respectively.
    The corresponding annihilation and creation operators satisfy the fundamental (anti)commutation relations $\{\hat f_j,\hat f_{j'}\}=0=[\hat b_j,\hat b_{j'}]$ and $\{\hat f_j,\hat f_{j'}^\dag\}=\delta_{j,j'}=[\hat b_j,\hat b_{j'}^\dag]$ for $j,j'\in\{1,2\}$.
    Furthermore, we find a diagonalization of the Hamiltonians via the eigenmodes
    \begin{equation}
    	\hat f_\pm=\frac{\hat f_1\pm\hat f_2}{\sqrt2}
    	\quad\text{and}\quad
    	\hat b_\pm=\frac{\hat b_1\pm\hat b_2}{\sqrt2},
    \end{equation}
    which remain fermionic and bosonic, respectively.

    For fermions, we find the Fock-basis expansion of $N$-particle states as
    \begin{equation}
    \begin{aligned}
    	|0,0\rangle_\pm
    	={}&
    	|0,0\rangle_\mathrm{FF}
    	=
    	|\mathrm{vac}\rangle,
    	\\
    	|1,0\rangle_\pm
    	={}&
    	\frac{
            |1,0\rangle_\mathrm{FF}+|0,1\rangle_\mathrm{FF}
    	}{\sqrt2},
    	\\
    	|0,1\rangle_\pm
    	={}&
    	\frac{
            |1,0\rangle_\mathrm{FF}-|0,1\rangle_\mathrm{FF}
    	}{\sqrt2},
    	\\
    	|1,1\rangle_\pm
    	={}&
    	|1,1\rangle_\mathrm{FF},
    \end{aligned}
    \end{equation}
    containing at most one excitation per mode because of the exclusion principle---likewise, the nilpotent operators $\hat f_j^{\dag 2}=0$.
    In addition, for bosons, we find the orthonormal $N$-particle states
    \begin{equation}
    	|m,n\rangle_\pm
    	=
    	\frac{\hat b_+^{\dag m}\hat b_-^{\dag n}}{\sqrt{m!n!}}|\mathrm{vac}\rangle
    	=\sum_{s=0}^N q_{s|m,n}|s,N-s\rangle_\mathrm{BB},
    \end{equation}
    with
    \begin{equation}
    \begin{aligned}
        q_{s|m,n}
        ={}&\sqrt{\frac{m!n!s!(N-s)!}{2^N}}
        \\
        {}& \times
        \sum_{l=\max\{0,s-m\}}^{\min\{s,n\}}
        \frac{(-1)^{n-l}}{l!(n-l)!(s-l)!(l+m-k)!}
    \end{aligned}
    \end{equation}
    and $m,n\in\mathbb N$.

    The vacuum states are in all cases identical.
    Also, the single-particle states, $(|0,1\rangle_{XY}\pm|1,0\rangle_{XY})/\sqrt2$, do not show differences for independent modes (i.e., $XY\in\{\mathrm{FF},\mathrm{BB}\}$) and dependent degrees of freedom, $XY=\mathrm{FB}$.
    Interesting differences appear for $N>1$.
    For instance, the two-particle states for two-mode fermions are always of a product form, regardless of the mode decomposition.
    Thus, in the following, we continue our comparison of only quasiparticles and bosons for $N=2$.

\begin{figure*}
	\includegraphics[width=.3\textwidth,trim={1.5cm 1.5cm 1.5cm 1.5cm}]{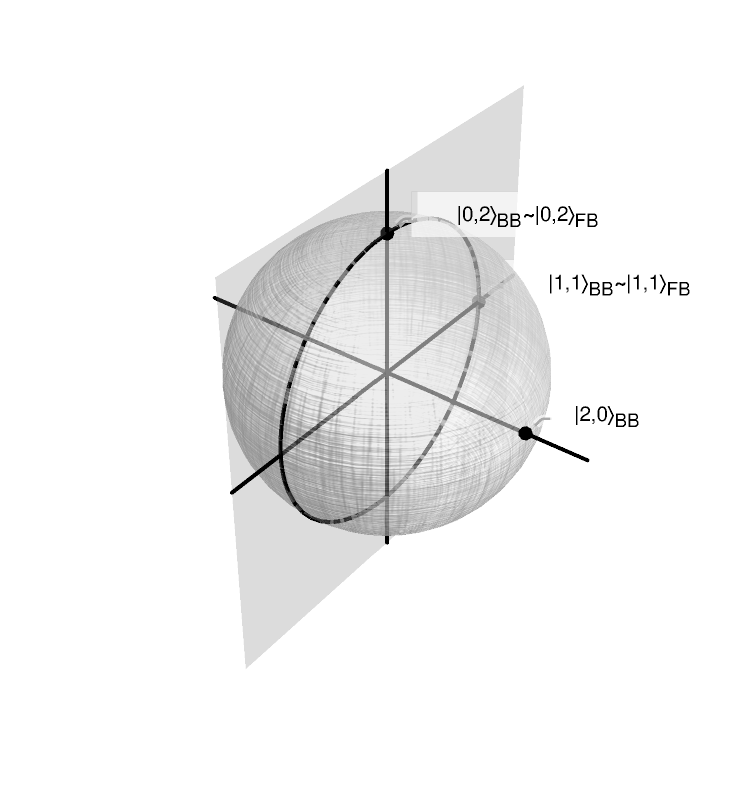}
	\includegraphics[width=.3\textwidth,trim={1.5cm 1.5cm 1.5cm 1.5cm}]{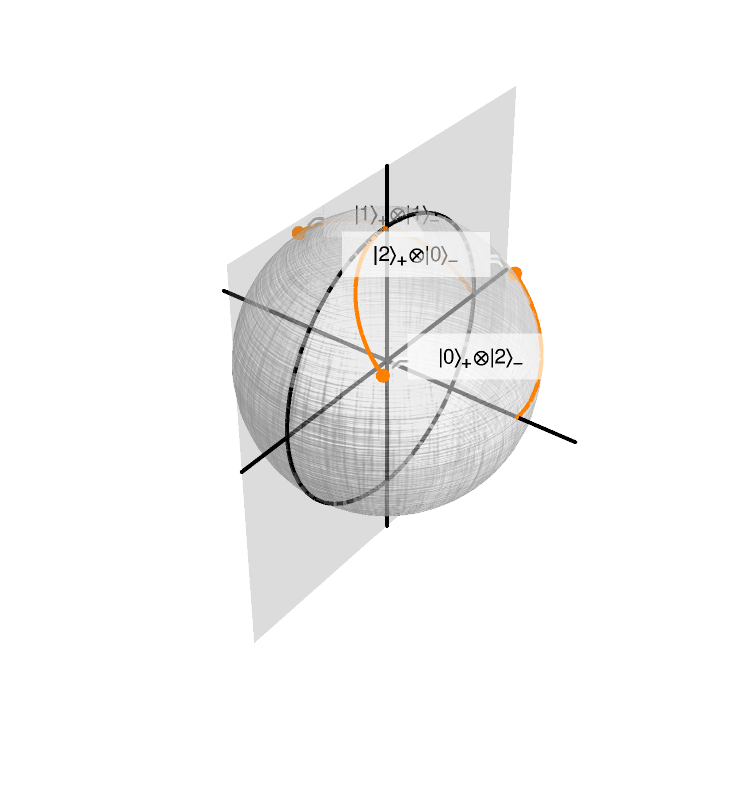}
	\includegraphics[width=.3\textwidth,trim={1.5cm 1.5cm 1.5cm 1.5cm}]{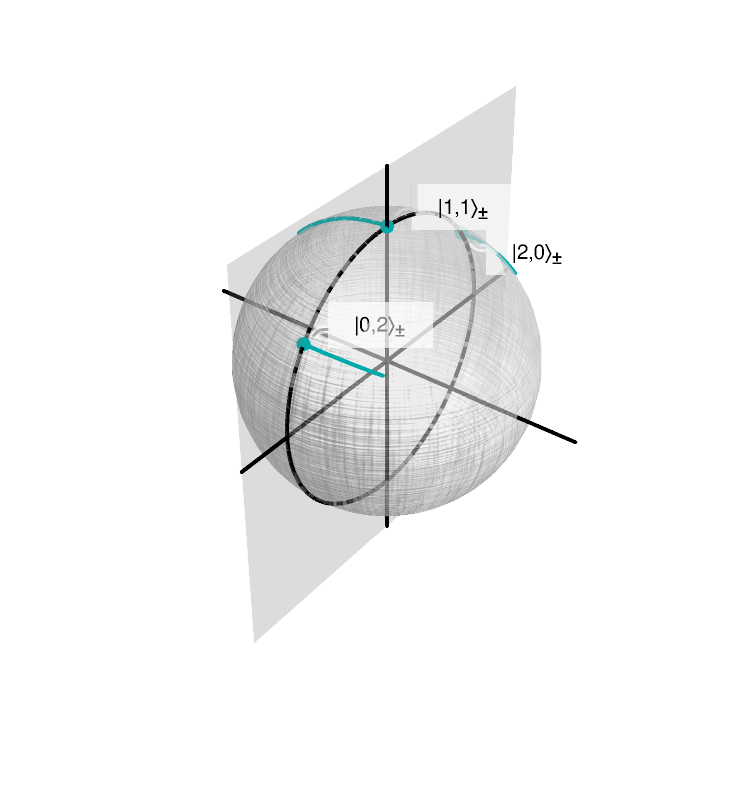}
	\caption{%
        Comparison of excitations of two bosons and two quasiparticles.
        The left plot shows the original space spanned by two bosons ($\mathrm{BB}$) and the subspace of the fermion-boson system ($\mathrm{FB}$).
        The center plot shows the effect of the rotation to the superposition modes, resulting in tensor-product basis states in the purely bosonic case.
        The right plot shows the projection $\hat P$, while preserving normalization, to the constrained quasiparticle states.
        In the middle plot, all states on the sphere but $|2-n\rangle_+\otimes|n\rangle_-$ and $-(|2-n\rangle_+\otimes|n\rangle_-)$ for $n\in\{0,1,2\}$ exhibit entanglement of independent degrees of freedom.
        However, for dependent degrees of freedom, every state on the circle except for the projected points $|2-n,n\rangle_\mathrm{\pm}$, together with their reflections, are the entangled ones in the right plot.
	}\label{fig:compare}
\end{figure*}

\subsection{Two bosons, quasiparticles, and physical projections}

    The purely bosonic case yields the $\pm$-product states, with the explicit expansion
    \begin{equation}
    \begin{aligned}
    	|2\rangle_+\otimes|0\rangle_-
        ={}&
        \frac{1}{2}
        \left(
            |2,0\rangle_\mathrm{BB}
            +\sqrt2|2,0\rangle_\mathrm{BB}
            +|0,2\rangle_\mathrm{BB}
        \right),
        \\
    	|1\rangle_+\otimes|1\rangle_-
        ={}&
        \frac{1}{\sqrt 2}
        \left(
            |2,0\rangle_\mathrm{BB}
            -|0,2\rangle_\mathrm{BB}
        \right),
        \\
    	|0\rangle_+\otimes|2\rangle_-
        ={}&
        \frac{1}{2}
        \left(
            |2,0\rangle_\mathrm{BB}
            -\sqrt2|2,0\rangle_\mathrm{BB}
            +|0,2\rangle_\mathrm{BB}
        \right).
    \end{aligned}
    \end{equation}
    For comparison, the explicit expansion of all pure factorizable two-quasiparticle states reads
    \begin{equation}
    \begin{aligned}
    	|2,0\rangle_\pm
    	={}&
    	\frac{1}{\sqrt 3}
    	\left(
            \sqrt{2}|1,1\rangle_\mathrm{FB}
            +|0,2\rangle_\mathrm{FB}
        \right),
        \\
        |1,1\rangle_\pm
        ={}&
        \left(
            -|0,2\rangle_\mathrm{FB}
        \right),
        \\
    	|0,2\rangle_\pm
    	={}&
    	\frac{1}{\sqrt 3}
    	\left(
            -\sqrt{2}|1,1\rangle_\mathrm{FB}
            +|0,2\rangle_\mathrm{FB}
        \right).
    \end{aligned}
    \end{equation}
    Up to a normalization, all instances of both cases are only distinguished through the presence ($\mathrm{BB}$ case) and the absence ($\mathrm{FB}$ case) of the $|2,0\rangle_{XY}$ component in the original mode decomposition.
    This makes sense as, for a single fermionic mode, we can identify
    \begin{equation}
    	\hat f=\hat P\hat b_1\hat P
    	\quad\text{and}\quad
    	\hat b=\hat b_2,
    \end{equation}
    with a projection operator in the Fock basis as
    \begin{equation}
        \label{eq:ProjectionFromFtoB}
    	\hat P=\sum_{n_1\in\{0,1\}}\sum_{n_2\in\mathbb N}|n_1,n_2\rangle_\mathrm{BB\,\,BB}\langle n_1,n_2|.
    \end{equation}
    One can straightforwardly check that a restriction of a bosonic annihilation operator to at most one excitation satisfies the properties of a fermionic annihilation operator.
    In other words, the above projection yields
    \begin{equation}
        \label{eq:BBprojectedToFB}
    	|m,n\rangle_\pm\propto\hat P|m\rangle_\mathrm{+}\otimes|n\rangle_-.
    \end{equation}
    Note that this relation holds true for $N=m+n>2$, too.
    \revision{Eventually, this presents the sought relation to the notion of dependent degrees of freedom, established in Sec. \ref{sec:DefinitionSection}.}

    The cases of independent degrees of freedom (two bosonic modes) and dependent degrees of freedom (quasiparticle modes) are related via a projection.
    This projection also explains the linear dependence since $\hat P$ is not bijective and maps an orthonormal tensor-product basis $|m\rangle_+\otimes|n\rangle_-$ to a linearly dependent set of vectors $|m,n\rangle_\pm$.
    In general, the definition of nonentangled states in Eq. \eqref{eq:classicalformPM} now takes the simpler form
    \begin{equation}
        \label{eq:classicalformPMprojective}
    	|x,y\rangle_\pm=\hat P\,|x\rangle_+\otimes|y\rangle_-,
    \end{equation}
    while choosing the scaling of the input states such that the projected result is properly normalized.
    In this form, the relation of nonentangled states $|x,y\rangle_\pm$ to tensor-product states, Eq. \eqref{eq:classicalformPMprojective}, becomes clearer.
    Also, the physical restriction of the composite system is achieved via the projection operator $\hat P$, projection onto the fermionic mode $\hat f$ as a part of both quasiparticle modes $\hat p_\pm$.
    See Fig. \ref{fig:compare} for a visualization.


\section{Multimode generalization}
\label{sec:Generalization}

%

    The framework discussed thus far is based on one fermionic and one bosonic mode.
    Beyond that, suppose we have $\hat f_j$, with $j\in\{1,\ldots,M\}$, and $\hat b_j$ as annihilation operators of our quantized modes in a $2M$-mode composite system.
    These operators obey the anticommutator and commutator relations,
    \begin{equation}
    \begin{aligned}
    	\{\hat f_j,\hat f_{j'}\}=0,
    	\quad&
    	[\hat b_j,\hat b_{j'}]=0,
    	\\
    	\{\hat f_j,\hat f_{j'}^\dag\}=\delta_{j,j'},
    	\quad\text{and}&\quad
    	[\hat b_j,\hat b_{j'}^\dag]=\delta_{j,j'}.
    \end{aligned}
    \end{equation}
    A linear mode transformation reads
    \begin{equation}
        \label{eq:InputOutputRelation}
    	\begin{bmatrix}
    		\hat p_{+,1}
    		\\
    		\vdots
    		\\
    		\hat p_{+,M}
    		\\
    		\hat p_{-,1}
    		\\
    		\vdots
    		\\
    		\hat p_{-,M}
    	\end{bmatrix}
    	=
    	\begin{bmatrix}
    		U_{+, F} & U_{+, B}
    		\\
    		U_{-, F} & U_{-, B}
    	\end{bmatrix}
    	\begin{bmatrix}
    		\hat f_{1}
    		\\
    		\vdots
    		\\
    		\hat f_{M}
    		\\
    		\hat b_{1}
    		\\
    		\vdots
    		\\
    		\hat b_{M}
    	\end{bmatrix},
    \end{equation}
    with matrices $U_{+, F}, U_{+, B},U_{-, F},U_{-, B}\in \mathbb C^{M\times M}$.

    Nonentangled states, akin to Eq. \eqref{eq:classicalformPM}, now are defined as
    \begin{equation}
    	|x^{(1)},\ldots,x^{(2M)}\rangle_\pm
    	=
    	\prod_{\substack{
            s\in\{+,-\},
            \\
            j\in\{1,\ldots,M\}
    	}}
    	\left(
            \sum_{m\in\mathbb N} \tilde x_{m}^{(s,j)}\hat p_{s,j}^{\dag m}
        \right)
    	|\mathrm{vac}\rangle.
    \end{equation}
    A convex hull construction then allows us to express mixed nonentangled (i.e., separable) states as statistical mixtures
    \begin{equation}
    \begin{aligned}
    	\hat\rho_\mathrm{sep.}
    	={}&
    	\int dP\left(x^{(1)},\ldots,x^{(2M)}\right)
    	\\
    	{}&\times
    	|x^{(1)},\ldots,x^{(2M)}\rangle_{\pm\,\,\pm}\langle x^{(1)},\ldots,x^{(2M)}|,
    \end{aligned}
    \end{equation}
    where $P$ denotes a joint probability distribution to account for classical correlations.
    Whenever a state $\hat\varrho$ is not in a separable configuration, $\hat\varrho\neq\hat\rho_\mathrm{sep.}$, we have an inseparable state that exhibits quantum correlations.

    Formally, the above definition of multimode nonentangled states for dependent degrees of freedom is identical to that for independent degrees of freedom.
    Hence, it is consistent with the existing approaches.
    However, as we have shown in Sec. \ref{sec:QuasiParticleEntanglement} for $M=2$, the entanglement and disentanglement of states can deviate from the common expectation.

    We also remark on the fact that the mapping in Eq. \eqref{eq:InputOutputRelation} does not have to be a unitary one.
    For instance, this allows us to describe product and entangled states in partially overlapping modes, e.g., $\hat p_{-,1}=\hat b_1$ and $\hat p_{-,2}=(\hat b_1+\hat b_2)/\sqrt2$, which is interesting in the context of partial coherence and entanglement in quantum optics \cite{SVPT15,KXLKW23}.


\section{Conclusion}
\label{sec:Conclusion}

    \revision{In summary, we established and explored entanglement for dependent degrees of freedom, which is compared to the archetypal notion of entanglement between independent degrees of freedom}.
    Defying common expectations, we showed that certain states are nonentangled in the case of dependent degrees of freedom which are entangled for independent notions of entanglement, and vice versa.

    The main physical systems to which we applied our approach were quasiparticle descriptions for light-matter quantum interfaces via a Jaynes--Cummings model.
    We showed that nonentangled quasiparticle states can be expressed through projections---accounting for physical constraints---that act on tensor-product states.
    The projection leads to linear dependencies, enabling us, for example, to construct NOON-type states which are not entangled, unlike their counterparts for independent degrees of freedom.
    Conversely, the true statement that eigenvectors of a local (and nondegenerate) Hamiltonian are necessarily factorizable is generally not true for dependent degrees of freedom.

    Furthermore, we compared our fermion-boson-composite quasiparticle description with purely bosonic and fermionic systems.
    For a single excitation, no distinction could be made between the different cases.
    However, we were able to identify unique entanglement characteristics for the different physical scenarios when considering at least two (quasi)particles.
    Eventually, a generalization to multipartite entanglement of dependent degrees of freedom, including the treatment of mixed states and physical systems other than quasiparticles, was established.

    Therefore, a framework was put forward to characterize quantum correlations between dependent degrees of freedom.
    This allows us to go beyond, for instance, the assumed well-separated nature of sender and receiver in quantum communication to accommodate for physical constraints not contributing to a quantum protocol as a quantum resource.

\begin{acknowledgments}
	The authors acknowledge funding through the Deutsche Forschungsgemeinschaft (DFG, German Research Foundation) via the Transregional Collaborative Research Center TRR 142 (Projects No. A04 and No. C10, Grant No. 231447078).
	This work was further supported through the Ministerium f\"ur Kultur und Wissenschaft des Landes Nordrhein-Westfalen through the project PhoQC: Photonisches Quantencomputing.
\end{acknowledgments}

\end{document}